\documentclass[amsmath,amssymb,12pt,superscriptaddress,nofootinbib]{revtex4}

\usepackage[dvipdfmx]{graphicx}
\usepackage{dcolumn}
\usepackage{bm}
\usepackage{color}
\usepackage{enumitem}
  \usepackage{tabularx}
   \newcolumntype{C}{>{\centering\arraybackslash}X}
   \newcolumntype{L}{>{\raggedright\arraybackslash}X}
   \newcolumntype{R}{>{\raggedleft\arraybackslash}X}
\setlistdepth{10}

\usepackage[normalem]{ulem}

\newcommand{\ii}{\mathrm{i}}
\newcommand{\dd}{\mathrm{d}}

\newcommand{\del}{\partial}
\newcommand{\ee}{{\rm e}}

\newcommand{\ellb}{{\ell_{\rm b}}}

\definecolor{DarkBlue}{rgb}{0,0.5,1} 

\definecolor{DarkRed}{rgb}{0.65,0,0}

\begin{document}
\baselineskip5.5mm

{\baselineskip0pt
\small
\leftline{\baselineskip16pt\sl\vbox to0pt{
                             \vss}}
\rightline{\baselineskip16pt\rm\vbox to20pt{
\vspace{-1.5cm}
            \hbox{YITP-23-157}
\vss}}
}

\vskip-1.cm

\author{Yusuke~Sakurai}

\affiliation{
\fontsize{12pt}{1pt}\selectfont
Division of Particle and Astrophysical Science,
Graduate School of Science, \\Nagoya University, 
Nagoya 464-8602, Japan
\vspace{1.5mm}
}

\author{Chul-Moon~Yoo}\email{yoo.chulmoon.k6@f.mail.nagoya-u.ac.jp}

\affiliation{
\fontsize{12pt}{1pt}\selectfont
Division of Particle and Astrophysical Science,
Graduate School of Science, \\Nagoya University, 
Nagoya 464-8602, Japan
\vspace{1.5mm}
}

\author{Atsushi~Naruko}

\affiliation{
\fontsize{12pt}{1pt}\selectfont
 Center for Gravitational Physics and Quantum Information, Yukawa Institute for Theoretical Physics, Kyoto University, Kyoto 606-8502, Japan
\vspace{1.5mm}
}

\author{Daisuke~Yamauchi}

\affiliation{
\fontsize{12pt}{1pt}\selectfont
Department of Physics, Faculty of Science,
Okayama University of Science, 1-1 Ridaicho, Okayama, 700-0005, Japan
\vspace{1.5mm}
}

\title{Axion Cloud Decay due to the Axion-photon Conversion 
\\with Multi-pole Background Magnetic Fields}

\vspace{1cm}


\begin{abstract}
\baselineskip5.5mm 

We consider axion cloud decay due to the axion-photon conversion 
with multi-pole background magnetic fields. 
We focus on the $\ell=m=1$ and $n=2$ mode for the axion field configuration 
since it has the largest growth rate associated with superradiant instability. 
Under the existence of a background multi-pole magnetic field, 
the axion field can be converted into the electromagnetic field 
through the axion-photon coupling. 
Then the decay rate due to the dissipation of the converted photons 
is calculated in a successive approximation. 
We found that the decay rate is significantly dependent on the 
azimuthal quantum number characterizing the background magnetic field, 
and can be comparable to or larger than the growth rate of the superradiant instability. 
\end{abstract}


\maketitle
\thispagestyle{empty}
\pagebreak

\section{Introduction}
Recently, axion-like particles (ALPs)\cite{Svrcek:2006yi,Arvanitaki:2009fg} have been actively 
investigated in many fields of science, such as particle physics, cosmology and astrophysics. 
The axion was originally introduced to solve the strong CP problem in quantum chromodynamics (QCD)\cite{Peccei:1977hh,Peccei:1977ur,Weinberg:1977ma,Wilczek:1977pj}. 
Observational and experimental constraints on the existence of ALPs including the QCD axion have been 
actively updated (see, e.g., Ref.~\cite{Zyla:2020zbs}). 

Let us focus on the 
black hole--axion cloud system, sometimes dubbed as the black hole (BH) atom. 
If the Compton wavelength of the axion is comparable to 
the mass of the spinning black hole, due to superradiant instability~\cite{Zouros:1979iw,Detweiler:1980uk,Brito:2015oca}, 
the angular momentum of the spinning black hole can be efficiently extracted. 
Then the axion field surrounding the black hole can be amplified and form a cloud. 
Possible observational signals from BH atoms~\cite{Cardoso:2011xi,Yoshino:2012kn,Yoshino:2013ofa,Brito:2014wla,Yoshino:2015nsa,Arvanitaki:2014wva,Brito:2017zvb,Baumann:2018vus,Zhang:2018kib,Zhang:2019eid,Baumann:2019ztm,Ding:2020bnl,Baumann:2021fkf,Roy:2021uye,Baumann:2022pkl,Chen:2022nbb,Tomaselli:2023ysb} and 
the final fate of this instability with or without external factors~\cite{Omiya:2020vji,Omiya:2022gwu,Omiya:2022mwv,Takahashi:2021yhy,Takahashi:2023flk,Spieksma:2023vwl} 
have been actively discussed. 

In general, an ALP may have the axion--photon coupling through the Chern-Simons term in the action. 
Many axion search experiments and observations give constraints on the magnitude of the coupling constant 
in each relevant mass scale~\cite{Anastassopoulos:2017ftl,Boutan:2018uoc,Ouellet:2018beu,Calore:2020tjw,Salemi:2021gck}. 
The birefringence due to this coupling term may cause observable imprints in cosmological and astrophysical observations~\cite{Carroll:1989vb,Carroll:1991zs,Harari:1992ea,Ivanov:2018byi,Fujita:2018zaj,Liu:2019brz,Fedderke:2019ajk,Caputo:2019tms,Chen:2019fsq,Yuan:2020xui,Basu:2020gsy}. 
In this paper, we consider the decaying process of the axion cloud through the axion--photon coupling. 

In Ref.~\cite{Yoo:2021kyv}, the authors calculated the decay rate of the axion cloud through the axion--photon coupling 
considering the dominant mode of superradiant instability given by $\ell=|m|=1$ and $n=2$, where $\ell$, $m$ and $n$ are the azimuthal, magnetic and principal quantum numbers. 
There, the typical size of the cloud has been assumed to be much larger than the horizon radius of the black hole, and the coupling constant 
has been assumed to be perturbatively small. 
In addition, in Ref.~\cite{Yoo:2021kyv}, the configuration of the background magnetic field is restricted to those of monopole and uniform configurations. 
In this paper, we adopt the same settings but different configurations of the background magnetic field given by 
the multi-pole magnetic field characterized by the azimuthal quantum number $\ellb$. 
Then we investigate the dependence of the decay rate on the configuration of the background magnetic field, or more specifically, on the value of $\ellb$. 


Throughout this paper, we use the natural units in which both 
the speed of light and the reduced Planck constant are unity, $c=\hbar=1$, 
and the gravitational constant is denoted by $G$.

\section{Equations of motion, background and perturbations}
\label{sec:mono}

Let us consider the axion-electro-magnetic system given by the following action:
\begin{equation}
S=\int\sqrt{-g}\dd^4x\left(-\frac{1}{4}F_{\mu\nu}F^{\mu\nu}-\frac{1}{4}\kappa \phi F_{\mu\nu}\tilde F^{\mu\nu}-\frac{1}{2}\nabla_\mu\phi \nabla^\mu \phi
-\frac{1}{2}\mu^2\phi^2\right), 
\end{equation}
where
\begin{equation}
\tilde F^{\mu\nu}=\frac{1}{2}\varepsilon^{\mu\nu\lambda\rho}F_{\lambda\rho}
\end{equation}
with $\varepsilon$ being the Levi-Civita tensor, and we neglected the non-linear self-interaction of the axion field. 
From the variation with respect to the axion field, we obtain the following equation of motion for the axion:
\begin{equation}
\left(\nabla_\mu\nabla^\mu-\mu^2\right)\phi=\frac{1}{4}\kappa F_{\mu\nu}\tilde F^{\mu\nu}. 
\label{eq:axioneom}
\end{equation}
The equations of motion for the gauge field are given by 
\begin{equation}
\nabla_\mu F^{\mu\nu}=-\kappa\tilde F^{\mu\nu}\nabla_\mu \phi, 
\label{eq:Maxwell}
\end{equation}
where we have used the following identity:  
\begin{equation}
\nabla_\mu\tilde F^{\mu\nu}=\frac{1}{2}\varepsilon^{\mu\nu\rho\lambda} \nabla_{\mu} F_{\rho\lambda}=0. 
\end{equation}
For the background geometry, we consider the Schwarzschild metric with the mass $M$ given by 
\begin{equation}
\dd s^2=-f(r)\dd t^2+\frac{\dd r^2}{f(r)}+r^2(\dd \theta^2+\sin^2\theta \dd \varphi^2), 
\end{equation}
where $f(r)=1-2GM/r$. 

In the following sections, we will consider 
the axion $\phi$ and the gauge fields $A^{\rm tot}_\mu$ given in the following form: 
\begin{eqnarray}
\phi&=&\delta \phi,\\
A^{\rm tot}_\mu&=&A^{\rm bg}_\mu +\delta A_\mu, 
\end{eqnarray}
where $A^{\rm bg}$ is the background gauge field satisfying Eq.~\eqref{eq:Maxwell} with $\phi=0$, 
and $\delta \phi$ and $\delta A_\mu$ are perturbations. 
We will consider the equations of motion for $\delta \phi$ and $\delta A_\mu$ 
at the linear order. 

In the form of the vector spherical harmonics~\cite{Zerilli:1974ai} and the Fourier mode expansion with the frequency $\omega$, 
we can expand the axion field and each component of the gauge field as 
\begin{eqnarray}
\delta\phi&=&\sum_{\ell\,  m} \Phi^{\ell\,  m} Y_{\ell\,  m}\ee^{\ii \omega t}, \\
\delta A_t&=&-\ii\sum_{\ell\,  m} A^{\ell\, m}_a Y_{\ell\,  m}\ee^{\ii \omega t}, 
\label{eq:AtY}
\\
\delta A_r&=&\sum_{\ell\,  m} A^{\ell\,  m}_b Y_{\ell\,  m}\ee^{\ii \omega t}, 
\label{eq:ArY}\\
\delta A_\theta&=&\sum_{\ell\,  m} \frac{1}{\sqrt{\ell (\ell +1)}}\left(A^{\ell \, m}_c\del_\theta Y_{\ell\,  m}+A^{\ell\,  m}_d
\frac{1}{\sin\theta}\del_\varphi Y_{\ell\,  m}\right)\ee^{\ii \omega t},
\label{eq:AthY}\\
\delta A_\varphi&=&\sum_{\ell\,  m} \frac{1}{\sqrt{\ell (\ell +1)}}\left(A^{\ell\,  m}_c\del_\varphi Y_{\ell\,  m}-A^{\ell\,  m}_d\sin\theta \del_\theta Y_{\ell\,  m}\right)\ee^{\ii \omega t}, 
\label{eq:AphY}
\end{eqnarray}
where, $Y_{\ell\,  m}$ is the spherical harmonic function of degree $\ell $ and order $m$, 
and $\Phi_{\ell\,  m}$, $A^{\ell\,  m}_a$, $A^{\ell\,  m}_b$, $A^{\ell\,  m}_c$ and $A^{\ell\,  m}_d$ are functions of $r$. 
The signs of $A^a_{\ell\,  m}$, $A^b_{\ell\,  m}$ and $A^c_{\ell\,  m}$ change as 
$(-1)^\ell $ for the parity transformation $(\theta, \varphi)\rightarrow(-\theta, \varphi+\pi)$ 
and the sign of $A^d_{\ell \, m}$ changes as $(-1)^{\ell +1}$, and they are called even and odd parity modes, respectively.



\section{Analysis with the multipole magnetic field}
\label{sec:mono}

Let us consider the following background gauge field: 
\begin{eqnarray}
A^{\rm bg}_\mu&=&pr^{-\ell_{\rm b}}\sin\theta\partial_\theta Y_{\ellb \, 0}({\rm d}\varphi)_\mu. 
\label{eq:Abg}
\end{eqnarray}
Although this multipole configuration is not exactly the solution of Eq.~\eqref{eq:Maxwell} with $\phi=0$,  
this can be an approximate solution under the condition $GM/r\ll 1$. 
In order to analytically evaluate the equations, hereafter, we impose the following conditions: 
\begin{eqnarray}
  &1/(GM)\gg\omega\sim\mu\gg1/r,
  \label{eq:approx}
  \\
  &p\kappa\ll a_0^{\ellb+1}, 
  \label{eq:smallcharge}
  \end{eqnarray}
where $a_0$, which corresponds to the Bohr radius, 
is defined by 
\begin{equation}
  a_0=\frac{1}{GM\mu^2}.  
\end{equation}
The gauge condition for the gauge field is taken as $A^c_{\ell\,  m}=0$ following Ref.~\cite{Zerilli:1974ai}. 

Since the background magnetic field violates the spherical symmetry, 
multiple modes are coupled with each other. 
Therefore the scalar field equation is given with the summation symbol, that is, 
\begin{equation}
\sum_{\ell\,  m}\left\{\left[\frac{1}{r^2}\del_r\left(r^2\del_r \Phi^{\ell\,  m}\right)+\omega^2\Phi^{\ell\,  m}
-\left(1-\frac{2GM}{r}\right)\mu^2\Phi^{\ell\,  m}-\frac{\ell (\ell +1)}{r^2}\Phi^{\ell\,  m}\right]Y_{\ell\,  m}\right\}=\frac{1}{4}\kappa F_{\mu\nu}\tilde F^{\mu\nu}. 
\label{eq:fullphieom_uni}
\end{equation}
Since we solve the equation in a successive approximation assuming the perturbatively small value of the coupling, 
the leading order equation is given by 
\begin{equation}
  \frac{1}{r^2}\del_r\left(r^2\del_r \Phi_0^{\ell\,  m}\right)+\left(\omega_0^2-\mu^2\right)\Phi_0^{\ell \, m}
  +\frac{2}{a_0r}\Phi_0^{\ell \, m}-\frac{\ell (\ell +1)}{r^2}\Phi_0^{\ell \, m}=0,  
  \label{eq:leadingphieom}
\end{equation}
where $\omega_0$ is the leading order term of $\omega$. 
This equation is equivalent to that for the Hydrogen atom. 
Then we suppose that the leading order is purely given by the mode with $\ell =m=1$ and the principal quantum number $n=2$. 
That is, we assume 
\begin{equation}
  \Phi_0:=\Phi_0^{1\, 1}=\omega_0\left(\frac{1}{2a_0\omega_0}\right)^{3/2}\frac{r}{\sqrt{3}a_0}\exp \left(-\frac{r}{2a_0}\right), 
  \label{eq:phi21}
  \end{equation}
  and the frequency is given by 
  \begin{equation}
  \omega_0^2-\mu^2=-\frac{1}{4a_0^2}, 
  \end{equation}
  where we have normalized $\Phi_0$ as $\omega_0\int \dd r r^2 \Phi_0^2=1$ keeping the dimension of $\Phi_0$ as the mass dimension one.

The equation for $\Phi^{1\,1}_1$ at the next order is given by
\begin{equation}
  \frac{d^2 \Phi_1^{1\,1}}{d r^2}+\frac{2}{r}\frac{d \Phi_1^{1\,1}}{d r}+\left(\omega_0^2-\mu^2\right)\Phi_1^{1\,1}
  +2\omega_0\omega_1\Phi_0^{1\,1}
  +\frac{2}{a_0r}\Phi_1^{1\,1}-\frac{\ell(\ell+1)}{r^2}\Phi_1^{1\,1}=\frac{1}{4}\left[\kappa F\tilde F\right]^{1\,1}=:S^{1\,1},  
  \label{eq:scl1st}
\end{equation}
where the superscript $^{1\,1}$ denotes the component of the mode given by $\ell=1$ and $m=1$, and 
the source term on the right-hand side will be explicitly evaluated later. 

Let us evaluate the leading order term on the right-hand side of Eq.~\eqref{eq:Maxwell}. 
The leading order can be given by substituting Eq.~\eqref{eq:Abg} and Eq.~\eqref{eq:phi21} into $\tilde F_{\mu\nu}$ and $\phi$ on the right-hand side 
of Eq.~\eqref{eq:Maxwell}, respectively. 
After some manipulation, the master equations for the even and odd parity modes are given by 
\begin{eqnarray}
&&r^2 \frac{\dd^2}{\dd r^2}A^{\ell\,  m}_b+(r^2\omega^2-\ell(\ell+1))A^{\ell \, m}_b
\cr
&&\hspace{1cm}=-\frac{\ii \kappa \ellb \omega_0 p r^{-\ellb}}{\ell(\ell+1)}\Bigl\{
r\Phi_0'\left[\ellb(\ellb+1)Y_{\ellb\,0} Y_{1\,1}-\partial_\theta Y_{\ellb\,0} \partial_\theta Y_{1\,1}\right]
\cr
&&\hspace{1cm}
+\Phi_0(\ellb+1)\left[(\ell(\ell+1)-\ellb(\ellb+1)) Y_{\ellb\,0} Y_{1\,1}
+\partial_\theta Y_{\ellb\,0} \partial_\theta Y_{1\,1}\right]
\Bigr\}^{\ell m}
\label{eq:evenA}
\end{eqnarray}
and 
\begin{eqnarray}
r^2 \frac{\dd^2}{\dd r^2}A^{\ell \, m}_d+(r^2\omega^2-\ell(\ell+1))A^{\ell \, m}_d
&
=&-\frac{\ii \kappa \ellb \omega_0 p r^{-\ellb+1}}{\sqrt{\ell(\ell+1)}}\Phi_0\Bigl\{
\frac{1}{\sin\theta}\partial_\theta Y_{\ellb\,0} \partial_\varphi Y_{1\,1}\Bigr\}^{\ell\, m}, 
\label{eq:oddA}
\end{eqnarray}
respectively.

Applying the formulae \eqref{eq:f1} -- \eqref{eq:f5} to the right-hand side of Eqs.~\eqref{eq:evenA} and \eqref{eq:oddA}, 
we can find that the 3 modes: $A^{\ellb\pm1\, 1}_b$ and $A^{\ellb\, 1}_d$ can be induced. 
Defining $\mathcal B_\pm:=A^{\ellb\pm1\, 1}_b/(\omega_0^{\ellb+1}r)$ and $\mathcal D:=A^{\ellb\, 1}_d/(\omega_0^{\ellb}r)$, 
by means of the Green's function method as in Ref.~\cite{Yoo:2021kyv}, we obtain the solutions for Eqs.~\eqref{eq:evenA} and \eqref{eq:oddA} 
with the regular center and outgoing boundary conditions. 
Then we obtain 
\begin{eqnarray}
{\rm Re}\mathcal B_+
&=&\kappa p \omega_0\sqrt{\frac{3}{8\pi}}\sqrt{\frac{\ellb+2}{(\ellb+1)(2\ellb+1)(2\ellb+3)}}\ellb 
j_{\ellb+1}(\omega_0 r)\cr
&&\times\int^\infty_0 \dd\xi j_{\ellb+1}(\xi)\xi^{-\ellb-1}\left[(\ellb+1)\Phi_0(\xi)+\ellb\xi\frac{\dd}{\dd\xi}\Phi_0\right], 
\label{eq:Bp}
\\
{\rm Re}\mathcal B_-
&=&\kappa p \omega_0\sqrt{\frac{3}{8\pi}}\sqrt{\frac{\ellb(\ellb-1)}{(2\ellb+1)(2\ellb-1)}}(\ellb+1) 
j_{\ellb-1}(\omega_0r)\cr
&&\times\int^\infty_0 \dd\xi j_{\ellb-1}(\xi)\xi^{-\ellb-1}\left[\Phi_0(\xi)-\xi\frac{\dd}{\dd\xi}\Phi_0\right], 
\label{eq:Bm}
\\
{\rm Im}\mathcal D
&=&-\kappa p \omega_0\sqrt{\frac{3}{8\pi}}\ellb 
j_{\ellb}(\omega_0r)
\int^\infty_0 \dd\xi j_{\ellb}(\xi)\xi^{-\ellb}\Phi_0(\xi), 
\label{eq:D}
\end{eqnarray}
where $j_\ell$ is the spherical Bessel functions, which give homogeneous solutions for Eqs.~\eqref{eq:evenA} and \eqref{eq:oddA}. 

Next, let us evaluate the source term of the scalar field equation \eqref{eq:scl1st}. 
At the leading order, we find 
\begin{equation}
  F_{\mu\nu}\tilde F^{\mu\nu}= \frac{1}{2}\varepsilon^{\mu\nu\lambda\rho}F_{\mu\nu}F_{\lambda\rho}\simeq
  \varepsilon^{\mu\nu\lambda\rho}F_{\mu\nu}^{\rm bg} F_{\lambda\rho}^{\rm \delta A}, 
\end{equation}
where $F^{\rm bg}_{\mu\nu}$ and $F^{\delta A}_{\mu\nu}$ are the field strengths calculated from 
the background gauge field \eqref{eq:Abg} and perturbative solution given in Eqs.~\eqref{eq:Bp} -- \eqref{eq:D} through 
Eqs.~\eqref{eq:AtY} -- \eqref{eq:AphY}. 
Then, after some manipulation, we obtain the following expression for $S^{\ell\,  m}$: 
\begin{eqnarray}
S^{1\, 1}&=&\Biggl\{\sum_\pm\Bigl[\ii \frac{\kappa p}{\omega_0}(\ellb\pm 1)(\ellb\pm 1+1)(\ellb+1)\ellb r^{-\ellb-4}A_b^{\ellb\pm1} 
Y_{\ellb\pm1\,1}Y_{\ellb\, 0}\cr
&&-\ii\frac{\kappa p}{\omega_0}\ellb r^{-\ellb-3}\frac{\dd}{\dd r}A^{\ellb\pm1}_b\partial_\theta Y_{\ellb\,0}\partial_\theta Y_{\ellb\pm1\,1}\Bigr]\cr
&&+\frac{\kappa p \omega_0 \ellb}{\sqrt{\ellb(\ellb+1)}}r^{-\ellb-3}A^{\ellb}_d\frac{1}{\sin\theta}\partial_\theta Y_{\ellb \,0}Y_{\ellb\, 1}\Biggr\}^{1\, 1}
\cr
&&+({\rm purely~real~ terms~associated~with~}\Phi_0~{\rm and}~\frac{\dd}{\dd r}\Phi_0), 
\end{eqnarray}
where we have used the equations of motion to remove $A_a$, $\frac{\dd}{\dd r}A_a$ and $\frac{\dd^2}{\dd r^2}A_b$. 

Then, using the formulae \eqref{eq:f6} -- \eqref{eq:f11}, we find 
\begin{eqnarray}
{\rm Im} S^{1\, 1}&=&-\sqrt{\frac{3}{8\pi}}\frac{\kappa p}{\omega_0}r^{-\ellb-4}\Biggl\{
\ellb(\ell_b+2)\sqrt{\frac{(\ellb+2)(\ellb+1)}{(2\ellb+3)(2\ellb+1)}}
\cr
&&\times{\rm Re}\left[-(\ellb+1)^2
A_b^{\ellb+1\, 1}+\ellb 
r\frac{\dd}{\dd r}A_b^{\ellb+1\, 1}\right]
\cr
&&+\ellb(\ellb+1)(\ellb-1)\sqrt{\frac{\ellb(\ellb-1)}{(2\ellb+1)(2\ellb-1)}}{\rm Re}\left[\ellb A_b^{\ellb-1\, 1}-r\frac{\dd}{\dd r}A_b^{\ellb-1\, 1}\right]\Biggr\}
\cr
&&-\sqrt{\frac{3}{8\pi}}\omega_0\kappa pr^{-\ellb-3}\ellb {\rm Im}A_d^{\ellb\, 1}, 
\end{eqnarray}
so that
\begin{eqnarray}
&&\int \dd r r^2 \Phi_0 {\rm Im}S^{1\, 1}=
\sqrt{\frac{3}{8\pi}}\kappa p\omega_0^{\ellb}\Biggl\{
\ellb(\ellb+2)\sqrt{\frac{(\ellb+2)(\ellb+1)}{(2\ellb+3)(2\ellb+1)}}
\cr
&&\hspace{1cm}\times 
\int\dd rr^{-\ellb-1}\left[(\ellb+1)\Phi_0+\ellb r \frac{\dd}{\dd r}\Phi_0\right] {\rm Re}\mathcal B_+\cr
&&\hspace{1cm}+\ellb(\ellb+1)(\ellb-1)\sqrt{\frac{\ellb(\ellb-1)}{(2\ellb+1)(2\ellb-1)}}
\int\dd rr^{-\ellb-1}\left[\Phi_0-r\frac{\dd}{\dd r}\Phi_0\right]{\rm Re}\mathcal B_-
\cr
&&\hspace{1cm}
-\omega_0\ellb\int \dd r r^{-\ellb} \Phi_0 {\rm Im}\mathcal D\Biggr\}. 
\end{eqnarray}
Taking the integral of the imaginally part of Eq.~\eqref{eq:scl1st} multiplied by $r^2\Phi_0/2$, we obtain the following result:
\begin{eqnarray}
{\rm Im} \omega_1&=&\frac{1}{2}\int \dd r r^2 \Phi_0 {\rm Im}S^{11}=\frac{3}{16\pi}\kappa^2p^2\omega_0^{2\ellb+3}\cr
&\times&\Biggl\{
\frac{\ellb^2(\ellb+2)^2}{(2\ellb+3)(2\ellb+1)}
I_{\ellb+1}^2
+\frac{\ellb^2(\ellb+1)^2(\ellb-1)^2}{(2\ellb-1)(2\ellb+1)}
I_{\ellb-1}^2
+
\ellb^2
I_{\ellb}^2
\Biggr\}>0, 
\end{eqnarray}
where
\begin{eqnarray}
I_{\ellb+1}&:=&\frac{1}{\omega_0}\int^\infty_0\dd\xi j_{\ellb+1}(\xi)\xi^{-\ellb-1}\left((\ellb+1)\Phi_0+\ellb\xi\frac{\dd\Phi_0}{\dd \xi}\right)
\cr
&\simeq&\sqrt{\frac{\pi}{3}}\frac{2^{-\ellb-3/2}}{\Gamma\left(\ellb+\frac{1}{2}\right)}(a_0\omega_0)^{-5/2}, 
\\
I_{\ellb-1}&:=&\frac{1}{\omega_0}\int^\infty_0\dd\xi j_{\ellb-1}(\xi)\xi^{-\ellb-1}\left(\Phi_0-\xi\frac{\dd\Phi_0}{\dd\xi}\right)
\cr&\simeq&\frac{\pi}{\sqrt{3}}\frac{2^{-\ellb-5/2}}{\Gamma\left(\ellb\right)}(a_0\omega_0)^{-7/2}, 
\\
I_\ellb&:=&\frac{1}{\omega_0}\int^\infty_0\dd\xi j_\ellb(\xi)\xi^{-\ellb}\Phi_0
\cr
&\simeq&\sqrt{\frac{\pi}{3}}\frac{2^{-\ellb-3/2}}{\Gamma\left(\ellb+\frac{1}{2}\right)}(a_0\omega_0)^{-5/2}\simeq I_{\ellb+1}.   
\end{eqnarray}
Here, $\Gamma$ is the Gamma function, and we evaluated the integrals in the limit $a_0\omega_0\rightarrow \infty$ in the last expressions. 
In the same limit, the imaginary part of $\omega_1$ is given by 
\begin{equation}
{\rm Im} \omega_1\simeq \kappa^2p^22^{-2\ellb-11}\ellb^2(\ellb+1)(2\ellb+1)(2\ellb+3)(5\ellb+7)\left[\Gamma\left(\ellb+\frac{5}{2}\right)\right]^{-2}a_0^{-5}\omega_0^{2\ellb-2}. 
\end{equation}
Introducing the 
value of the magnetic field $B$ around $r\sim a_0$ as 
\begin{equation}
B:=pa_0^{-\ellb-2}, 
\end{equation}
we obtain
\begin{equation}
{\rm Im} \omega_1\simeq \kappa^2B^22^{-2\ellb-11}\ellb^2(\ellb+1)(2\ellb+1)(2\ellb+3)(5\ellb+7)\left[\Gamma\left(\ellb+\frac{5}{2}\right)\right]^{-2}a_0^{2\ellb-1}\omega_0^{2\ellb-2}. 
\end{equation}

\section{Comparison with the growth rate of the superradiant instability}
\label{sec:comp}

\begin{figure}[htbp]
\begin{center}
\includegraphics[scale=0.8]{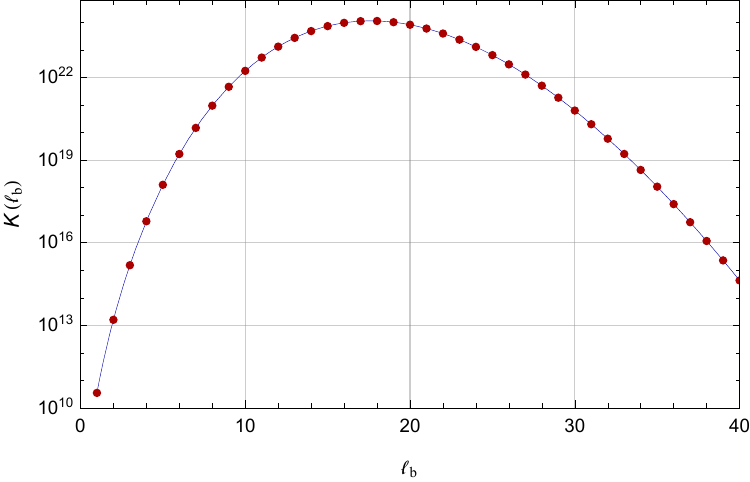}
\caption{The value of $K(\ellb)$ as a function of $\ellb$. 
}
\label{fig:Klb}
\end{center}
\end{figure}

We compare this value with the growth rate of the superradiant instability given by 
\begin{equation}
\omega_{\rm sr}\simeq \frac{1}{48}\left(a_0\omega_0\right)^{-8}\omega_0. 
\end{equation}
The ratio can be evaluated as 
\begin{eqnarray}
\frac{{\rm Im}\omega_1}{\omega_{\rm sr}}&\simeq&
3 \kappa^2B^22^{-2\ellb-7}\ellb^2(\ellb+1)(2\ellb+1)(2\ellb+3)(5\ellb+7)\left[\Gamma\left(\ellb+\frac{5}{2}\right)\right]^{-2}a_0^{2\ellb+7}\omega_0^{2\ellb+5}
\cr&\simeq&
K(\ellb)\left(\frac{\kappa}{10^{-12}{\rm GeV^{-1}}}\right)^2\left(\frac{B}{10^3{\rm G}}\right)^2
\left(\frac{\mu}{10^{-18}{\rm eV}}\right)^{-2\ellb-9}\left(\frac{M}{4\times 10^6 M_\odot}\right)^{-2\ellb-7}
\cr&\simeq&10^{-10}
K(\ellb)\left(\frac{\kappa}{10^{-12}{\rm GeV^{-1}}}\right)^2\left(\frac{B}{10^3{\rm G}}\right)^2
\left(\frac{\mu}{10^{-13}{\rm eV}}\right)^{-2\ellb-9}\left(\frac{M}{40 M_\odot}\right)^{-2\ellb-7}, 
\end{eqnarray}
where 
\begin{equation}
K(\ellb)=3.9\times 10^6\times 2^{2\ellb}3^{-2\ellb}5^{4\ellb}\ellb^2(\ellb+1)(2\ellb+1)(2\ellb+3)(5\ellb+7)\left[\Gamma\left(\ellb+\frac{5}{2}\right)\right]^{-2}. 
\label{eq:Kellb}
\end{equation}
The value of $K(\ellb)$ is shown as a function of $\ellb$ in Fig.~\ref{fig:Klb}. 
The value of $K(\ellb)$ takes the maximum value $K(\ellb)\sim 1.1\times 10^{24}$ at $\ellb=18$. 
Fig.~\ref{fig:Klb} shows that the significance of the decay due to the axion-photon conversion 
highly dependent on the configuration of the background magnetic field. 

\section{summary and discussion}
We derived the decay rate of the axion cloud around a black hole due to the axion-photon conversion 
under the existence of a multipole background magnetic field. 
While the axion cloud can grow due to the superradiant instability extracting the rotation energy of the black hole, 
the axion can be converted to photons and dissipate to infinity. 
The decay rate ${\rm Im} \omega_1$ is roughly given by the form ${\rm Im} \omega_1\sim  K(\ellb) \kappa^2 B^2 (GM)^{-2\ellb+1} \mu^{-2\ellb}$, 
where $\kappa$, $\ellb$, $M$, $\mu$ and $B$ are the coupling constant between the axion and photon, 
the azimuthal quantum number for the multipole magnetic field, 
the mass of the black hole and the mass of the axion, 
the typical value of the magnetic field around at the radius $\sim 1/(GM\mu^2)$, 
respectively, and the $\ellb$ dependent prefactor $K(\ellb)$ is given by Eq.~\eqref{eq:Kellb} in the text. 
We found that the prefactor $K(\ellb)$ is highly dependent on the 
value of $\ellb$ and it takes the maximum value about $10^{24}$ at $\ellb=18$. 
Therefore we conclude that the efficiency of the decay due to the axion-photon conversion 
heavily depends on the configuration of the background magnetic field. 
We have assumed several conditions to perform the analytic evaluation (see Eqs.~\eqref{eq:approx}
and \eqref{eq:smallcharge} in the text). 
Nevertheless, for some specific parameter regions, the decay rate can be comparable to the growth rate 
of the superradiant instability. 
Thus extensions to more realistic situations around astrophysical black holes 
would be important issues, which we leave for future work.

\section*{Acknowledgements}
This work was supported by JSPS KAKENHI Grant
Numbers JP20H05850~(C.Y.), JP20H05853~(C.Y.), 20H05852~(A.N.), JP22K03627~(D.Y.), JP19H01891~(A.N. and D.Y.), JP23H01171~(A.N. and D.Y.). 
A.N. thanks to the molecule workshop "Revisiting cosmological non-linearities in the era of precision surveys" YITP-T-23-03 since discussions during the workshop were useful for this work. 
We acknowledge the hospitality at APCTP during the focus research program 
"Black Hole and Gravitational Waves: from modified theories of gravity to data analysis" 
where part of this work was done. 

\appendix

\section{Useful formulae}
Here we list useful formulae used in the text: 
\begin{eqnarray}
\del_\theta Y_{\ell\,m}&=&\frac{1}{2}\ee^{-\ii\varphi}\sqrt{(\ell-m)(\ell+m+1)}Y_{\ell\,m+1}
\cr&&\hspace{1cm}
-\frac{1}{2}\ee^{\ii\varphi}\sqrt{(\ell+m)(\ell-m+1)}Y_{\ell\,m-1}, 
\label{eq:f1}
\\
Y_{\ell_1\,m_1}Y_{\ell_2\,m_2}&=&\sum_{\ell}\left[\frac{(2\ell_1+1)(2\ell_2+1)(2\ell+1)}{4\pi}\right]^{1/2}
\cr&&\hspace{1cm}\times
\left(
\begin{array}{ccc}
\ell_1&\ell_2&\ell\\
m_1&m_2&-m_1-m_2
\end{array}
\right)
Y^*_{\ell~-m_1-m_2}
\left(
\begin{array}{ccc}
\ell_1&\ell_2&\ell\\
0&0&0
\end{array}
\right), 
\label{eq:f2}
\\
Y_{\ell \,0}Y_{1\,1}&=&\sqrt{\frac{3}{8\pi}}\Bigl(
\sqrt{\frac{(\ell+1)(\ell+2)}{(2\ell+1)(2\ell+3)}}
Y_{\ell+1\,1}
\cr&&\hspace{1cm}
-\sqrt{\frac{\ell(\ell-1)}{(2\ell+1)(2\ell-1)}}Y_{\ell-1\,1}\Bigr), 
\label{eq:f3}
\\
\partial_\theta Y_{\ell \,0}\partial_\theta Y_{1\,1}
&=&-\sqrt{\frac{3}{8\pi}}\Bigl(
\ell\sqrt{\frac{(\ell+1)(\ell+2)}{(2\ell+1)(2\ell+3)}}
Y_{\ell+1\,1}
\cr&&\hspace{1cm}
+(\ell+1)\sqrt{\frac{\ell(\ell-1)}{(2\ell+1)(2\ell-1)}}Y_{\ell-1\,1}\Bigr), 
\label{eq:f4}
\\
\frac{1}{\sin\theta}\partial_\theta Y_{\ell\,0}\partial_\varphi Y_{1\,1}&=&-\ii \sqrt{\frac{3}{8\pi}}\sqrt{\ell(\ell+1)}Y_{\ell\,1}, 
\label{eq:f5}
\\
\frac{1}{\sin\theta}\del_\theta Y_{\ell \,0}&=&\sqrt{\frac{2\ell+1}{4\pi}}\frac{1}{\sin\theta}\del_\theta P_{\ell}(\cos\theta)\cr
&=&-\sqrt{\frac{2\ell+1}{4\pi}}\sum^{[\frac{\ell-1}{2}]}_{k=0}\left[\left\{2(\ell-2k)-1\right\}P_{\ell-1-2k}(\cos\theta)\right]\cr
&=&-\sqrt{2\ell+1}\sum^{[\frac{\ell-1}{2}]}_{k=0}\left[\sqrt{2(\ell-2k)-1}Y_{\ell-1-2k\,0}\right], 
\label{eq:f6}
\\
\left[Y_{\ell+1\,1}Y_{\ell \,0}\right]^{1\,1}&=&\sqrt{\frac{3}{8\pi}}\sqrt{\frac{(\ell+1)(\ell+2)}{(2\ell+1)(2\ell+3)}}
\label{eq:f7}
\\
\left[Y_{\ell-1\,1}Y_{\ell \, 0}\right]^{1\,1}&=&-\sqrt{\frac{3}{8\pi}}
\sqrt{\frac{\ell(\ell-1)}{(2\ell+1)(2\ell-1)}},
\label{eq:f8}
\\
\left[\del_\theta Y_{\ell\,0}\del_\theta Y_{\ell+1\, 0}\right]^{1\,1}&=&\sqrt{\frac{3}{8\pi}}\ell(\ell+2)\sqrt{\frac{(\ell+1)(\ell+2)}{(2\ell+1)(2\ell+3)}}
\label{eq:f9}
\\
\left[\del_\theta Y_{\ell\,0}\del_\theta Y_{\ell-1\, 0}\right]^{1\,1}&=&-\sqrt{\frac{3}{8\pi}}(\ell+1)(\ell-1)
\sqrt{\frac{\ell(\ell-1)}{(2\ell+1)(2\ell-1)}},
\label{eq:f10}
\\
\left[\frac{1}{\sin\theta}\del_\theta Y_{\ell\,0}Y_{\ell\,1}\right]^{1\,1}&=&-\sqrt{\frac{3}{8\pi}}\sqrt{\ell(\ell+1)},  
\label{eq:f11}
\end{eqnarray}
where the $2\times 3$ matrices in \eqref{eq:f2} are the Wigner 3-$j$ symbols.


\end{document}